\documentclass[conference]{IEEEtran}
\IEEEoverridecommandlockouts

\usepackage{amsmath,graphicx}
\usepackage{cite}
\DeclareMathOperator*{\E}{\mathbb{E}}
\usepackage{amssymb,amsfonts}

\usepackage{algorithmic}
\usepackage{graphicx}
\usepackage{textcomp}
\usepackage{xcolor}
\usepackage{bm}
\usepackage{subcaption}
\usepackage{algorithmic}
\usepackage{epsfig}
\usepackage{float}
\usepackage{comment}
\usepackage{mathtools,amssymb}
\usepackage{tikz}
\usepackage{pgfplots}

\usetikzlibrary{pgfplots.groupplots}
\DeclareMathOperator{\vect}{vec}
\newcommand{\RNum}[1]{\uppercase\expandafter{\romannumeral #1\relax}}
\usepackage[linesnumbered,ruled,vlined]{algorithm2e}

\DeclareMathOperator*{\argminA}{arg\,min} 

   \definecolor{mycolor1}{rgb}{0.92900,0.69400,0.12500}%
\definecolor{mycolor2}{rgb}{0.49400,0.18400,0.55600}%
\definecolor{mycolor3}{rgb}{0.46600,0.67400,0.18800}%
\definecolor{mycolor4}{rgb}{0.30100,0.74500,0.93300}%
\definecolor{mycolor5}{rgb}{0.63500,0.07800,0.18400}%
\definecolor{mycolor6}{rgb}{0.00000,0.44700,0.74100}%
\def\BibTeX{{\rm B\kern-.05em{\sc i\kern-.025em b}\kern-.08em
    T\kern-.1667em\lower.7ex\hbox{E}\kern-.125emX}}

\begin{document}

\title{Supervised Learning Based Super-Resolution DoA Estimation Utilizing Antenna Array Extrapolation}

\author{\IEEEauthorblockN{Udaya Sampath K.P. Miriya Thanthrige, Aya Mostafa Ahmed and Aydin Sezgin\thanks{This work was funded by the Deutsche Forschungsgemeinschaft (DFG, German Research Foundation) – Project-ID 287022738 – TRR 196 (S02 and S03 Projects).}}
\IEEEauthorblockA{Institute of Digital Communication Systems,\\
Ruhr University Bochum, Germany.\\
Email: \{udaya.miriyathanthrige, aya.mostafaibrahimahmad, aydin.sezgin\}@rub.de}
}

\IEEEpubid{978-1-7281-5207-3/20/ \copyright\ 2020 IEEE. Personal use is permitted, but republication/redistribution requires IEEE permission.}
\maketitle

\begin{abstract}
In this paper, we propose a novel algorithm based on supervised learning
for antenna array extrapolation for the purpose of super resolution DoA estimation. We use multiple signal classification (MUSIC) as a DoA estimation technique to estimate the DoA. In order to reduce the computational burden, existing approaches focus on interpolating the missing elements in a virtual array using a sparse array (or non-uniform linear arrays) or employ antenna selection within the same aperture. In contrast, here we utilize an uniform linear array (ULA) using a low number of antennas (within a small array
aperture) to extrapolate another ULA with a higher number of antennas
(within a bigger aperture). This approach is not restricted to any
specific antenna array configuration, in fact it can be
generalized to any array configuration. We propose an algorithm which utilizes the advances of supervised learning (dictionary learning) to find a mapping between the receive signal of the small to the bigger ULA using multiple training scenarios. Each scenario is trained using the received signal of both apertures from multiple targets within different angle ranges in a multiple input multiple output (MIMO) radar setup. In the testing phase, however, we can only use the small ULA to approach the performance of the bigger ULA to a certain extent. For example, if the small ULA is a $10\times10$ antenna array and the bigger ULA is a $16\times16$ antenna array, then by means of dictionary learning, we predict the receive signal of the bigger ULA using only the small ULA. Simulations show that using our approach, the training based small ULA can resolve more targets, especially in low SNR environments when compared to the untrained ULA.
\end{abstract}

\begin{IEEEkeywords}
Sparse signal processing, dictionary
learning, DoA estimation, MIMO radar, MUSIC
\end{IEEEkeywords}

\section{Introduction }
\label{sec:intro}
Super resolution DoA estimation is a vital research area for many applications like radar and wireless communications. However, DoA estimation algorithms face a lot of challenges that affect it's precision, e.g., the existence of very close targets and low signal to noise ratio (SNR). This can be solved by increasing aperture size and the number of antennas, which enhances the DoA algorithm resolution \cite{Trees}. However, this increases the system complexity and computational cost.
 As a solution to this, several strategies were used such as non-uniform linear arrays (NLA) \cite{abramovich1999positive}, sparse NLA \cite{tuncer2007direction}, \cite{eldar} and co-prime arrays \cite{qin2015generalized}. In \cite{eldar}, the authors suggested a sparse non uniform array with low number of transmit/receive antennas placed randomly on a large aperture, which may be not suitable for some applications. Moreover, in \cite{tuncer2007direction} and \cite{liu2016coprime} array interpolation techniques were used on sparse NLA and co-prime arrays to interpolate the missing elements in the array geometry to achieve a desired virtual array using the same antenna aperture. 
Those methods assume the existence of large aperture size antenna array, where they select which antennas to operate (on/off) from this array to enhance the DoA estimation. In contrast, the main contribution of this paper is proposing a novel approach that uses the advances of supervised learning and sparse signal processing. This is done by learning the mapping between the received signal of a low antenna setup (small aperture) to the received signal of a high antenna setup (large aperture). Thus, by this mapping, performance similar to the high antenna setup using only the low antenna setup can be achieved. In our method both low and high antenna setups are only required during the training. After the training, only the low antenna setup is required to predict the received signal of the high antenna setup.\\ 
In this paper, our supervised learning method is based on data-driven coupled dictionary learning. Here, we learn a coupled dictionary pair which has a common sparse representation for the received signals of both high and low antenna setups. Afterwards, these learned dictionaries are used to predict the received signal of high antenna array setup using the received signal of low antenna setup. A dictionary is defined based on a sparse representation. The sparse representation is approximating a given signal as linear combination of few basis functions. A collection of these functions are called a dictionary \cite{aharon2006k}. For some signals predefined dictionaries such as wavelets and curvelets are suitable for sparse representation. However, learning a dictionary from data would result in an improved sparse representation \cite{aharon2006k}. Sparse representation and dictionary learning has been used for many applications such as signal separation \cite{7378306} and super resolution imaging \cite{5466111}.\\ \IEEEpubidadjcol
For verification, our method is applied on MIMO radar and we use MUSIC \cite{Music} to estimate the DoA of the targets using the predicted signal. Although, other DoA estimation methods such as compressing sensing based methods \cite{fortunati2014single}, ESPRIT \cite{duofang2008angle} can be applied to estimate the DoA. Moreover, our approach can be generalized to other array geometries such as NLA, sparse NLA or co-prime arrays i.e., learn the mapping between two co-prime arrays. As, the main goal of this paper is to learn the mapping between received signal of two given antenna arrays which have different aperture size. It is worth noting that in this work, we are not interested in antenna selection or antenna array design.
\section{System Model} \label{smodel}
\subsection{MIMO Radar Signal Model}\label{mimoradar}
In this section, we describe a general signal model for the
MIMO radar. Consider a co-located MIMO radar system with $M$ transmit antennas (TX), each transmitting a train of $P$ non overlapping pulses, and $N$ receive antennas (RX). The TX and RX antennas are placed in uniform linear array (ULA) with $d$ spacing between each antenna. Here, $d=\lambda_0/2$ and $\lambda_{0}$ is the operating wavelength. 
We assume that there are $K$ targets in the radar scene, each has direction of arrival (DoA) at angle $\theta_k$, with radar cross section (RCS) modeled by $\alpha_{k,p} \in \mathbb{C}$. Moreover, each antenna $m$ transmits narrow-band signal denoted by $s_m(t)$, where all $M$ transmitted signals are assumed to be perfectly orthogonal such that
\begin{equation}
\label{orth}
\int_{0}^{T} s_m(t)s_q(t)^{*} dt=\left\{\begin{matrix}
1 & m=q\\ 
0& m\neq q,
\end{matrix}\right. 
\end{equation}
where $()^{*}$ denotes a conjugate operator, and $T$ is the pulse repetition interval. Hence, the
signal received at target $k$ after transmitting the $p$-th pulse is given by $\sum_{m=1}^{M}\mathbf{a}_t^T(\theta_k) s_m(t-pT)$. Here, $\mathbf{a}_t^T(\theta_k)$ is the steering vector towards target $k$ and defined as
$
\mathbf{a}_t(\theta_k)=\left[1,e^{j{\rho}d \sin\theta_k},\hdots,e^{j{\rho}d (M-1)\sin\theta_k}\right]^T,$ where $\rho=\frac{2\pi}{\lambda_{0}}$. Let us define $\mathbf{r}(t)\in \mathbb{C}^{N}$ as the received signal, which is given by
\begin{equation}
\mathbf{r}(t)=\sum_{k=1}^{K}\sum_{p=0}^{P-1} \alpha_{k,p} \mathbf{a}_r(\theta_k)\mathbf{a}_t^T(\theta_k) 
\mathbf{s}(t-pT) + \mathbf{n}(t),
\end{equation}  
where $\mathbf{s(t)}=[s_1(t),\hdots,s_m(t)]^T$, and $\mathbf{a}_r(\theta_k)$ is the receive steering vector for target $k$, and it is defined as $
\mathbf{a}_r(\theta_k)=\left[1,e^{j{\rho}d \sin\theta_k},\hdots,e^{j{\rho}d (N-1)\sin\theta_k}\right]^T.$ Here, $\mathbf{n}(t) \in \mathbb{C}^{N}$ is independent and identically distributed (i.i.d) Gaussian noise with variance $\sigma^2$.
We assume that the target RCS is fixed during pulse interval $T$ and changes independently from one pulse to another, following the Swerling model \RNum{2} \cite{swerling}.
Afterwards, at each RX antenna, the received signal is cross correlated with filters matched to the transmitted wave-forms, such that 
\begin{gather}
\begin{split}
\mathbf{Z}_p(t)& =\int_{0}^{T} \mathbf{r}(t) \mathbf{s}^{H}(t-pT) dt, \\
& = \sum_{k=1}^{K}\sum_{p}^{P} \alpha_{k,p} \mathbf{a}_r(\theta_k)\mathbf{a}_t^T(\theta_k) \mathbf{I}
+\int_{0}^{T} \mathbf{n}(t)\mathbf{s}^{H}(t-pT) dt,
\end{split}
\label{eq:received_pulse}
\raisetag{40pt}
\end{gather}
where $\mathbf{I}$ is an identity matrix, as we assume perfect orthogonality between the transmitted signals as shown in eq. \eqref{orth}.
Furthermore, received signal $\mathbf{Y}$$\in$  $\mathbb{C}^{MN\times P}$ is defined as
\begin{equation}
\mathbf{Y}=\mathbf{A}(\theta)\mathbf{X}+\mathbf{N}.
\label{eq:MIMO_receved}
\end{equation}
Here, the matrix in eq. \eqref{eq:received_pulse} is converted into a column vector by vectorization, denoted as $\vect(\mathbf{Z}_p)$, and stacked in a matrix $\mathbf{Y}=[\vect(\mathbf{Z}_1),\hdots,\vect(\mathbf{Z}_P)]$. Here, $\mathbf{A}(\theta)$ contains the virtual array steering vector   $\mathbf{v}(\theta_k)= \mathbf{a}_t(\theta_k) \otimes \mathbf{a}_r(\theta_k),$ such that $\mathbf{A}(\theta)=[\mathbf{v}(\theta_1),\hdots,\mathbf{v}(\theta_K)].$
The RCS of $K$ targets is contained in the $K \times P$ matrix $
\mathbf{X}$, where $\mathbf{X}=[\mathbf{x}_1,\hdots,\mathbf{x}_P]$ and $
\mathbf{x}_p=[\alpha_{1,p},\hdots,\alpha_{K,p}]^T$.
In the next section, we are interested in finding a dictionary ($\bm{D}$ of size $MN \times L$) that represents $\mathbf{Y}$ in eq. \eqref{eq:MIMO_receved} in sparse manner (i.e., $\mathbf{Y}=\bm{D}\bm{W}+\mathbf{N})$.
 Here, $\bm{W}$ is the sparse coefficient matrix of size $L \times P$, which is column-wise sparse. This can be justified by considering a simple example, if we consider a single pulse ($P=1$), then $\mathbf{Y}$ is reduced to a vector of size $MN \times 1$. This vector can be represented in sparse manner with respect to a known dictionary or transform, as in reality, the number of targets $K$ is much smaller than $MN$. 
 \begin{figure}[!t]
	\centering
	\includegraphics[trim=0.0cm 0.0cm 0.0cm 0.0cm, clip,width=0.25\textwidth]{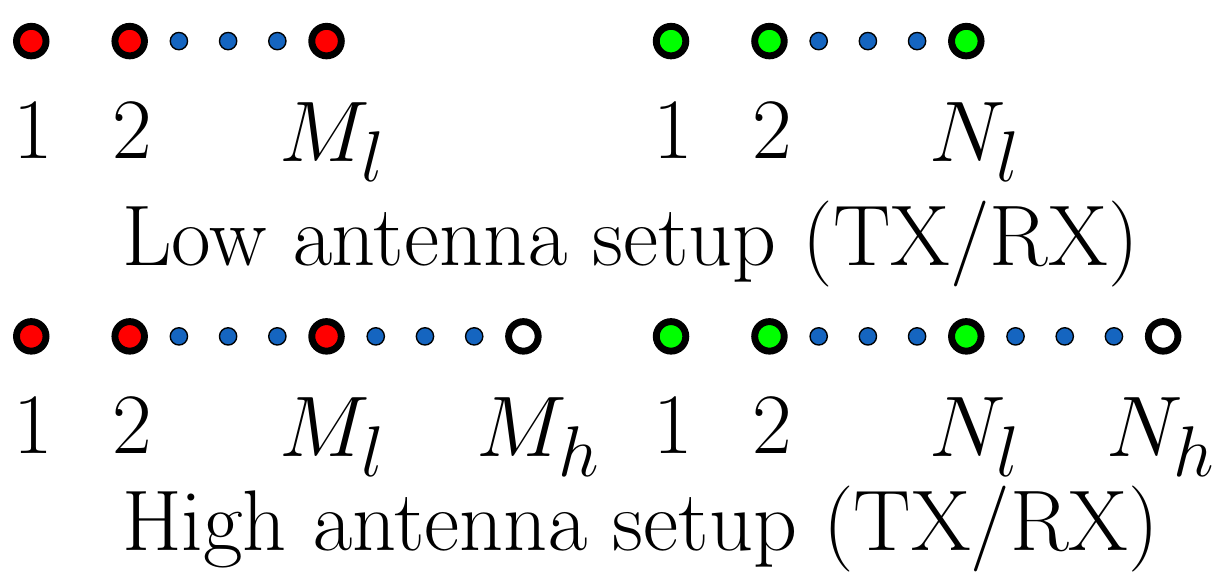}
	\caption{Description of low and high antenna setup}
	\label{ant_array}
\end{figure}
\subsection{Coupled Dictionary learning for MIMO radar}\label{dlearning}
In this section, we discuss the proposed dictionary learning based signal prediction for the MIMO radar setup. We consider two antenna setups referred to as high and low based on the number of antennas as shown in Fig. \ref{ant_array}. For low and high antenna setups, the number of TX and RX antennas are given by $M_l$, $N_l$, $M_h$ and $N_h$, respectively. Here, the low antenna setup is a subset of the high antenna setup as shown in Fig. \ref{ant_array}, where both share the first $M_l$ and $N_l$ elements for TX and RX, respectively. Now,
suppose that, the received signal (in eq. \eqref{eq:MIMO_receved}) for low and high antenna setups are given as $\bm{Y}_l$ and $\bm{Y}_h$, respectively. We assume that there exist a coupled dictionary pair ($\bm{D}_l$, $\bm{D}_h$) which has the same sparse representation for both $\bm{Y}_l$ and $\bm{Y}_h$. The dictionary pair ($\bm{D}_l$, $\bm{D}_h$) is used to learn the mapping between $\bm{Y}_l$ and $\bm{Y}_h$. The dimension of each dictionary is $M_lN_l \times L$ and $M_hN_h \times L$, respectively. Here, $L$ is the size of the dictionary (where $L \geq M_hN_h$). Thus, $\bm{Y}_l$ and $\bm{Y}_h$ can be decomposed as
\begin{equation}
	\begin{aligned}
		\bm{Y}_l &= \bm{D}_l \bm{W} + \bm{N}_l,\\
		\bm{Y}_h &= \bm{D}_h \bm{W} + \bm{N}_h.
	\end{aligned}
	\label{D_le_eq1}
\end{equation}
Note that, the received signals ($\bm{Y}_l$, $\bm{Y}_h$) are complex valued matrices. Therefore, in dictionary learning and signal prediction we treat real and imaginary components separately.
Here, $\bm{W}$ is the sparse coefficient matrix of size $L \times P$ which is column-wise sparse and $i$-th column of the matrix $\bm{W}$ is denoted by $\bm{w}_i$. The $\bm{N}_l$, $\bm{N}_h$ are the noise matrices, which are modeled as zero mean random Gaussian noise with covariance
matrix of $\sigma^2\mathbf{I}$. Based on \eqref{D_le_eq1}, the coupled dictionary learning problem can be formulated as
\begin{gather}
	\label{D_le_eq2}
	\begin{aligned}
		\left\{\bm{\hat{W}}, \bm{\hat{D}}_l, \bm{\hat{D}}_h\right\} &=\argminA_{\bm{W},  \bm{D}_l, \bm{D}_h}  \left\lVert \left[\begin{matrix}
			\bm{Y}_l \\
			\bm{Y}_h 
		\end{matrix} \right] - \left[\begin{matrix}
			\bm{D}_l \\
			\bm{D}_h 
		\end{matrix}\right] \bm{W} \right\rVert_F^2 \\
		& \ \ \  \ \text{s.t} \ \parallel \bm{w}_i \parallel_0 \ \le s, \forall i,
	\end{aligned}
	\raisetag{150pt}
	\end{gather}
 Here, $s$ is user defined sparsity constraint.\\ Now, $\bm{\tilde{Y}}$ $=$$ \left[ \begin{matrix}
\bm{Y}_l \\
\bm{Y}_h 
\end{matrix} \right]$ and $\bm{D}$$ =$$\left[\begin{matrix}
\bm{D}_l \\
\bm{D}_h 
\end{matrix}\right] $. The optimization problem in \eqref{D_le_eq2} is non-convex and thus challenging. By relaxing the $l^0$-norm with the $l^1$-norm, eq. \eqref{D_le_eq2} can be written  as 
 \begin{gather}
	\label{D_le_eq3}
	\begin{aligned}
		\left\{\bm{\hat{W}}, \bm{\hat{D}} \right\} &=\argminA_{\bm{W},  \bm{D}}  \parallel \bm{\tilde{Y}}  - \bm{D}   \bm{W} \parallel_F^2 \\
		& \ \ \  \ \text{s.t} \ \parallel \bm{w}_i \parallel_1 \ \le s, \forall i.
	\end{aligned}
	\raisetag{150pt}
\end{gather}
This problem in \eqref{D_le_eq3} is referred as dictionary learning \cite{naumova2018fast}. It can be solved numerically by alternating minimization between $\bm{W}$ and $\bm{D}$ \cite{aharon2006k}, \cite{mairal2009online}. In this paper, we choose the online dictionary learning (ODL) \cite{mairal2009online} approach to solve \eqref{D_le_eq3} due to its computational efficiency and accuracy. In \eqref{D_le_eq3}, the value of $s$ is selected from a predefined uniform range which provides the lowest reconstruction error $\left(\parallel \bm{\tilde{Y}}-\bm{D} \bm{W} \parallel^2_2 \right)$.
Next, we use $P$ number of pulses to generate the training data ($\bm{\acute{Y}}_l$, $\bm{\acute{Y}}_h$) from each setup.
\subsubsection{Training}
\label{sub_section:training}
In the training stage, dictionary learning is performed to learn the dictionary pair ($\bm{D}_l$, $\bm{D}_h$) using $\bm{\acute{Y}}_l$, $\bm{\acute{Y}}_h$. Due to the complexity, we observe that learning only one pair of dictionaries that capture very large angle range (i.e., $0$ to $90$ degrees) is often not enough. Therefore, to enhance the robustness, in this paper, we propose to divide the angle range in to several grids (i.e., $0:25$, $10:35$, ... , $65:90$ degrees) and then train several dictionary pairs based on the angle regions (grids). As a pre-processing step we normalize $\bm{\acute{Y}}_l$ and $\bm{\acute{Y}}_h$ column-wise with respect to $\bm{\acute{Y}}_l$ to bring the data to a common scale. This improves the prediction accuracy. Also, we subtract the column-wise mean from the signal, to make the signal to have zero mean column-wise. This is done to avoid the dictionary learning process to be ill-conditioned \cite{naumova2018fast}. The training process is repeated for a certain number of iteration till it converges (till the reconstruction error is within a predefined range).
\subsubsection{Prediction (Testing) }
 After learning the dictionary pairs ($\bm{D}_l$ and $\bm{D}_h$) for different angle grids in the training, for a given received signal of the low antenna setup $\bm{\tilde{Y}}_l$, the
received signal of high antenna setup $\bm{\hat{Y}}_h$ is predicted using algorithm \ref{D_algo:pre}. Here,
a dictionary pair must be selected based on the angle grid of the targets. To make that selection, we use the DoA estimation of the low antenna setup to find an initial guess of the angle range (grid). Based on this, the dictionary pair is selected. In algorithm \ref{D_algo:pre}, the least absolute shrinkage and selection operator (LASSO) is used to calculate the sparse coefficients \cite{tibshirani1996regression}. Here, $\lambda$ is a regularization parameter.
\begin{algorithm}[h]
	\LinesNumberedHidden
	\DontPrintSemicolon 
	\KwIn{ \hspace{5pt}\\  
		\hspace{5pt} Received signal of low antenna setup ($\bm{\tilde{Y}}_l$), $\bm{D}_l$,  $\bm{D}_h$}
	\textbf{Initialization:}\\
	\hspace{5pt} Data pre-processing\\
	\hspace{15pt} $\bm{\phi}$ $=$  column-wise mean of $\bm{\tilde{Y}}_l$\\
	\hspace{15pt} $\bm{\Phi}_l$$=$ reshape $\bm{\phi}$ to have same dimension of $\bm{\tilde{Y}}_l$\\
	\hspace{15pt} $\bm{\tilde{Y}}_l$$=$ $\bm{\tilde{Y}}_l$$-$ $\bm{\Phi}_l$\\
	\hspace{15pt} $\bm{\breve{Y}}_{l}$$=$ column-wise normalization of $\bm{\tilde{Y}}_l$ \\
	\hspace{5pt} Calculate the sparse coefficient matrix using $\bm{\breve{Y}}_{l}$ 
	\vspace{-5pt}
	\begin{gather*}
	\begin{aligned}
	\{\bm{\hat{W}}\} =\argminA_{\bm{W}} \parallel \bm{\breve{Y}}_{l} - \bm{D}_l \bm{W} {\parallel}^2_2 + \lambda {\parallel}\bm{W}{\parallel}_1.
	\end{aligned}
	\end{gather*}
	\hspace{5pt} \text{Initial prediction: } $\bm{\hat{Y}}_h =   \bm{D}_h \bm{\hat{W}}$.\\
	\Repeat{convergence }{
		Joint sparse coefficient update: \\
		$\bm{D}$ $=$ $\left[\begin{matrix}
		\bm{D}_l \\
		\bm{D}_h 
		\end{matrix}\right] $, $\bm{\breve{Y}} = \left[ \begin{matrix}
		\bm{\breve{Y}}_{l} \\
		\bm{\hat{Y}}_h 
		\end{matrix} \right]$, 
		\vspace{-5pt}
		\hspace{15pt} \begin{gather*}
		\begin{aligned}
		\{\bm{\hat{W}}\} &=\argminA_{\bm{W}} \parallel \bm{\breve{Y}} - \bm{D} \bm{W} {\parallel}^2_2 + \lambda {\parallel}\bm{W}{\parallel}_1.\\
		\text{Prediction: }
		\bm{\hat{Y}}_h &=  \bm{D}_h\bm{\hat{W}}.
		\end{aligned}
		\end{gather*}
		\vspace{-10pt}	
	}
	$\bm{\hat{Y}}_h$ $=$ column-wise de-normalization of $\bm{\hat{Y}}_h$ w.r.t  $\bm{\tilde{Y}}_l$. \\
	$\bm{\Phi}_h$ $=$ reshape $\bm{\phi}$ to have same dimension of $\bm{\hat{Y}}_h$.\\
	$\bm{\hat{Y}}_h$ $=$ $\bm{\hat{Y}}_h$$+$ $\bm{\Phi}_h$.\\
	\Return{\text{Predicted signal} $=$ $\bm{\hat{Y}}_h$.}\;
	\caption{{\sc}Signal prediction using coupled dictionary learning}
	\label{D_algo:pre}	
\end{algorithm}
\subsection{DoA Estimation using MUSIC}
\label{DOA}
In order to evaluate our prediction algorithm, we apply MUSIC on the received signal in \eqref{eq:MIMO_receved}, to estimate the DoA for all the targets. The resolution of MUSIC depends on the number of array elements. Thus, with a low antenna setup, it might not be able to resolve all the targets. Hence, we use MUSIC to evaluate the ability of the predicted receive signal to resolve targets, which could not be resolved in the low antenna setup. Note that, MUSIC depends on the orthogonality of the target eigen vector to the noise eigen vector \cite{Music}. To this end, the covariance matrix of the received signal in \eqref{eq:MIMO_receved} can be written as
\begin{equation}
\label{cov}
\begin{split}
\mathbf{R}&=\E{[\mathbf{Y}\mathbf{Y}^H}]  = \mathbf{A}(\theta)\E{[\mathbf{X}\mathbf{X}^H}]\mathbf{A}^H(\theta)+\sigma^2 \mathbf{I}\\
&=\mathbf{U}_x\mathbf{\Lambda}_x\mathbf{U}_x^H+\mathbf{U}_n\mathbf{\Lambda}_n\mathbf{U}_n^H,
\end{split}
\end{equation}
 where $\mathbf{U}_x$, $\mathbf{U}_n$ are matrices containing the eigen vectors, which represent the signal and noise subspace respectively. $\mathbf{\Lambda}_x = \mathrm{diag}(\lambda_1,\hdots,\lambda_K)$ and $\mathbf{\Lambda}_n=\mathrm{diag}(\lambda_{K+1},\hdots,\lambda_{MN})$ contain the corresponding eigen values of the target and the noise respectively. Hence, the expression of the MUSIC spectrum is given by $P_{MU}(\theta)=\left({\mathbf{v}^H(\theta)\mathbf{U}_n\mathbf{U}^H_n \mathbf{v}(\theta})\right)^{-1}$.
\begin{figure*}[!t]
	\centering
	\begin{subfigure}[t]{.33\textwidth}
		\centering
		\includegraphics[width=0.9\textwidth]{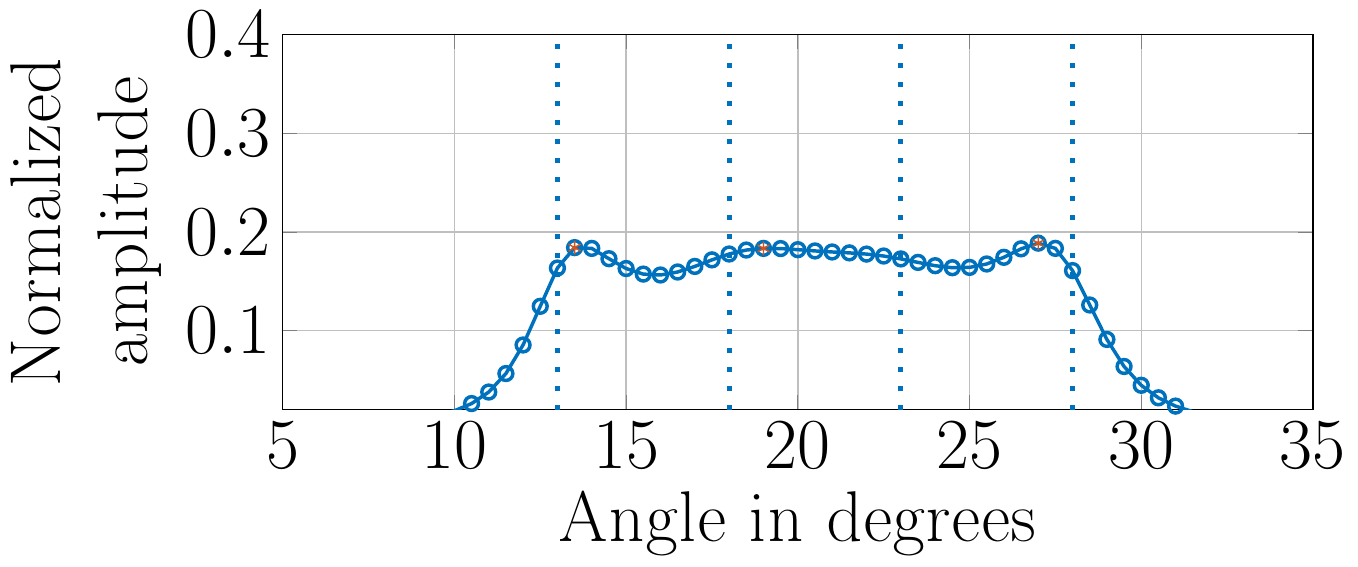}
		\caption{Low antenna setup ($10 \times 10$)}
		\label{sim_figure3_a}
	\end{subfigure}\hfill
	\begin{subfigure}[t]{.33\textwidth}
		\centering
		\includegraphics[width=0.9\textwidth]{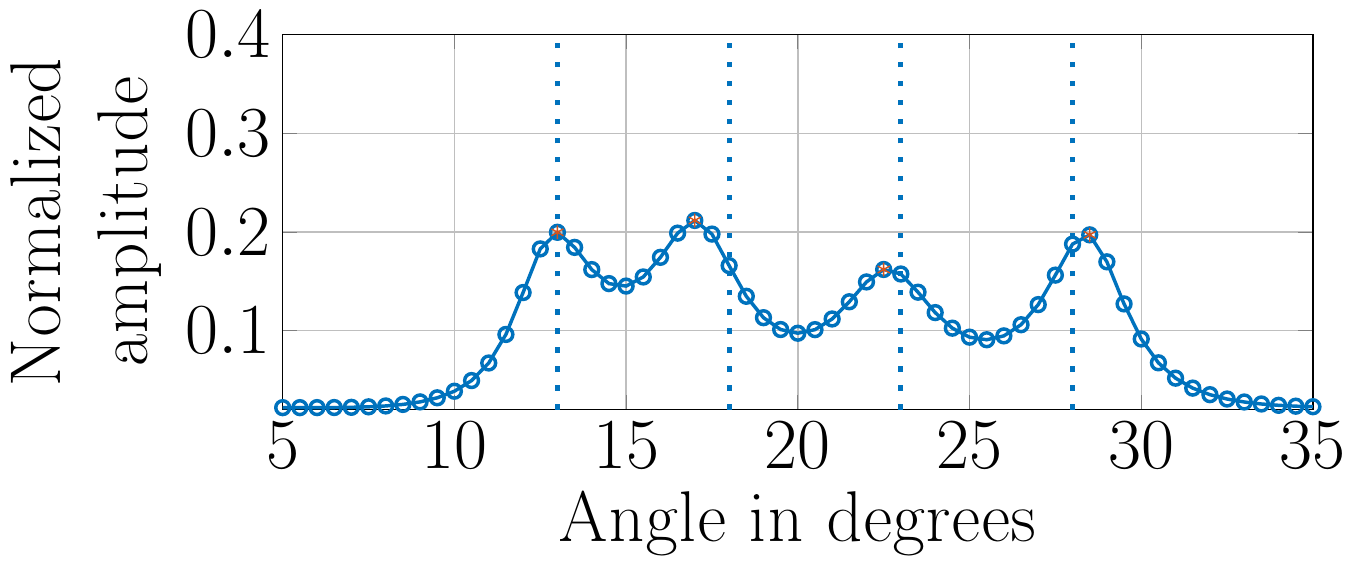}
		\caption{Predicted using $10 \times 10$ to $16 \times 16$}
		\label{sim_figure3_b}
	\end{subfigure}\hfill
	\begin{subfigure}[t]{.33\textwidth}
		\centering
		\includegraphics[width=0.9\textwidth]{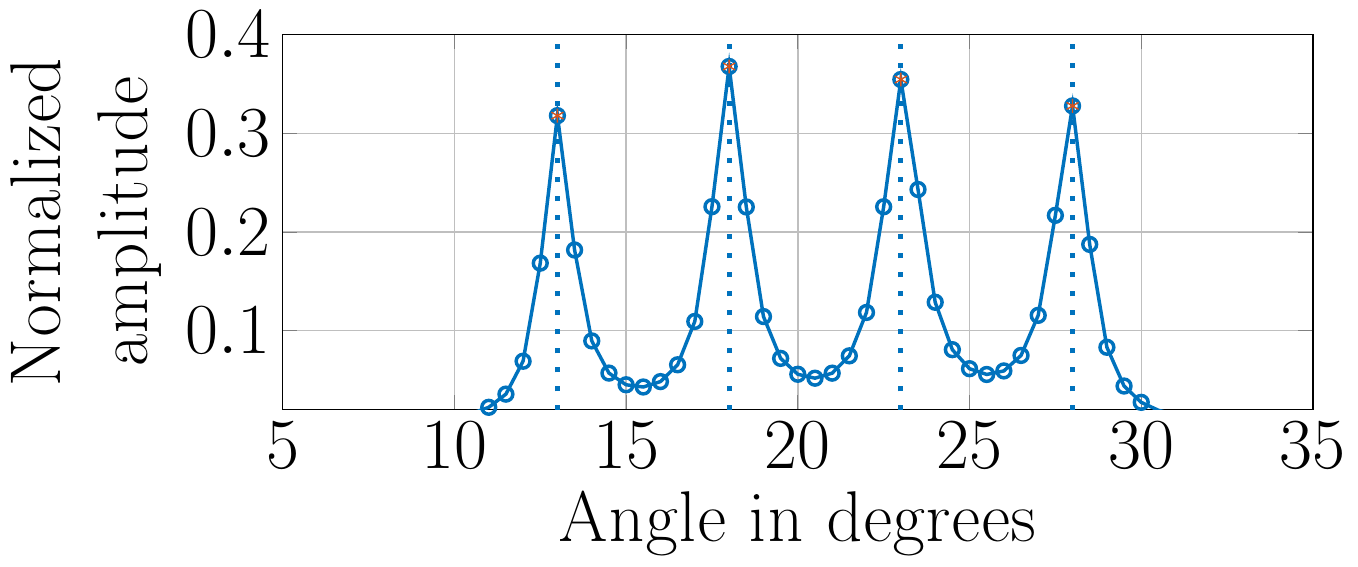}
		\caption{High antenna setup ($16 \times 16$)}
		\label{sim_figure3_c}
	\end{subfigure}
	\caption{MUSIC Spectrum for four targets ($13, 18, 23, 28$ degrees) at angle range $10$ to $35$ for SNR$=$$-10$ dB} \label{sim_figure3}
\end{figure*}
\begin{table}[!t]
	\centering
	\begin{tabular}{|l|l|l|l|}
		\hline
		$N_l$ $=$ $M_l$                                                                                 & $10$  & $N_h$ $=$  $M_h$                                                                                & $16$    \\ \hline
		
		$L$                                                                                    & $512$ & $\lambda$                                                                               & $0.01$  \\ \hline
		Angle grid 1  & $10:35$ &  Angle grid 2  & $20:45$ \\ \hline
		Angle grid 3  & $30:55$ &  Number of targets ($K$) & $4$ \\ \hline
	\end{tabular}
	\caption{Parameter for the simulation}
	\label{table_1}
		\vspace{-0.5cm}
\end{table}
\section{Simulation Setup and Results} \label{ssetup}
Here, multiple simulations were performed to evaluate our algorithm. We demonstrate the DoA estimation of a single test, then we simulate the average behavior using Monte Carlo simulations. It is vital to select the appropriate parameters for the dictionary learning such as $L$, regularization parameter ($\lambda$) and number of iterations in eq. \eqref{D_le_eq3}. For that purpose, we empirically set them by experience to minimize the reconstruction error $\left(\parallel \bm{\tilde{Y}} - \bm{D} \bm{W} \parallel^2_2 \right)$ in the training phase. In the training phase, the number iterations for dictionary learning is set as $300$ and number of training data samples $P$ in eq. \eqref{eq:MIMO_receved} is set as $45000$. Other parameters used for the simulations are listed in Table \ref{table_1}.
\subsection{DoA estimation example using algorithm \ref{D_algo:pre}} \label{one_instance}
Here, we test our algorithm for a single test case. For this case four targets ($K=4$) located within angle grid $1$ given in the Table \ref{table_1} are used. Here, the SNR is set to be $-10$ dB and $100$ snapshots are used. Fig. \ref{sim_figure3} shows the comparison of DOA for the four targets across different array setups. On one hand, fig. \ref{sim_figure3_a}, the DOA estimation is shown for the low antenna setup ($10\times10$), where it can be depicted that the low antenna setup is unable to resolve all targets. On the other hand, fig. \ref{sim_figure3_c} shows the DOA estimation for the high antenna setup ($16 \times 16$), where the targets are correctly resolved. Fig. \ref{sim_figure3_b} shows the DoA estimation of the predicted signal using the algorithm \ref{D_algo:pre}, in which we use the trained dictionaries which learn the mapping from the received signal of low antenna setup to the received signal of high antenna setup. It can be seen, that our algorithm can correctly predict all the angles, performing similar results to the ($16\times16$) antennas using only the ($10\times10$).\\
To further evaluate our algorithm, we conduct different scenarios through Monte Carlo simulations. Root mean square error (RMSE) of the estimated angles is used as a metric, which is defined as \textbf{$\sqrt{ \frac{1}{K}\sum_{k=1}^{K} (\theta_{e,k}-\theta_{a,k})^2}$}, where the estimated and actual angle of the $k$-th target is given as $\theta_{e,k}$ and $\theta_{a,k}$ respectively. In the training phase, the angles are generated randomly within the angle grids as shown in Table \ref{table_1}. However, in the testing phase, for a fair comparison we set the angle gap between adjacent targets to $5$ degrees.
\begin{figure}[!t]
	\centering  
		\includegraphics[width=0.4\textwidth]{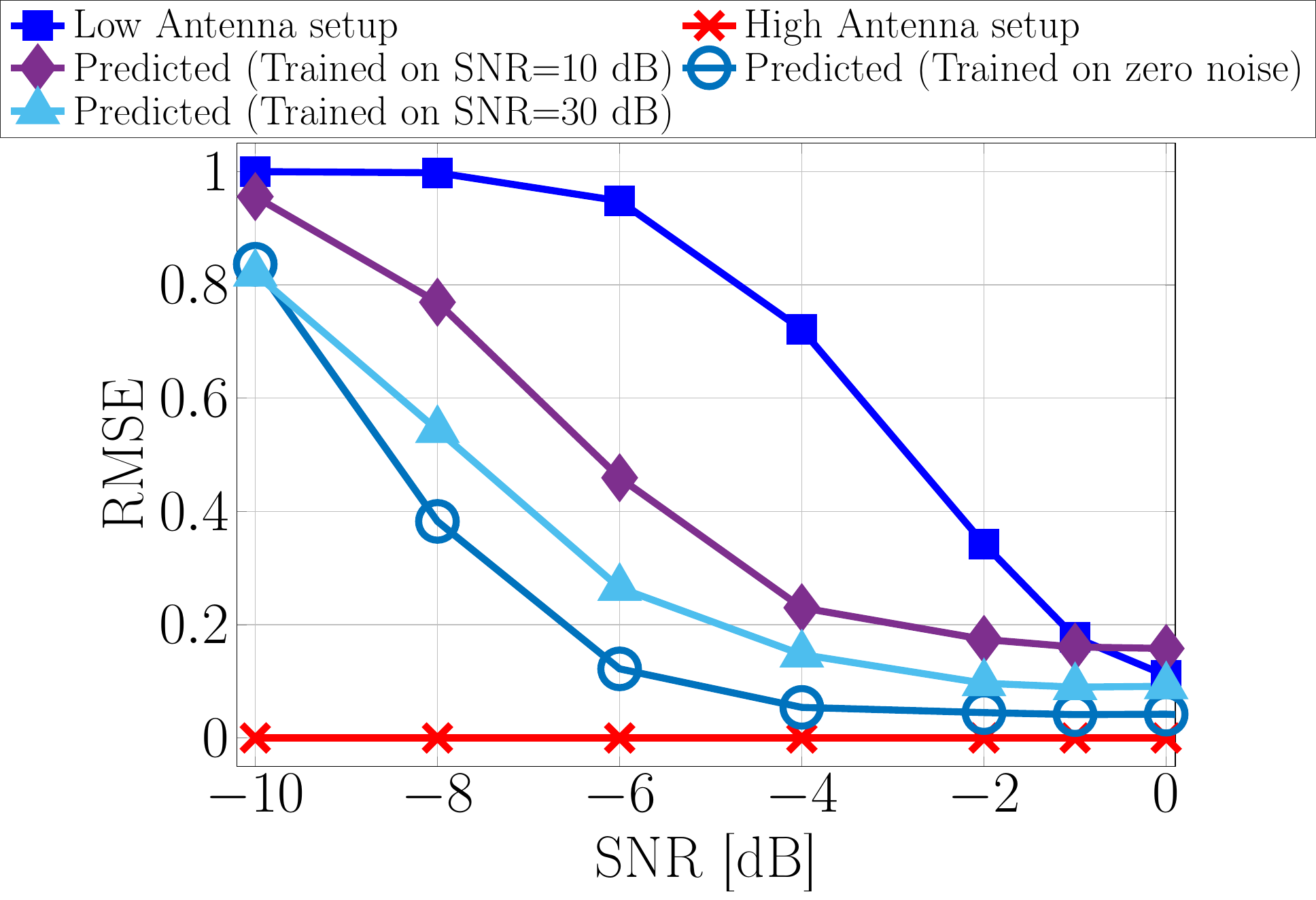}
		\caption{DoA estimation for predicted signal using dictionaries trained on different SNR for angle grid $2$}
		\label{sim_figure4}
		\vspace{-0.4cm}
	\end{figure}
\begin{figure*}[!t]
	\begin{minipage}[t]{0.6\linewidth}
		\includegraphics[width=\textwidth]{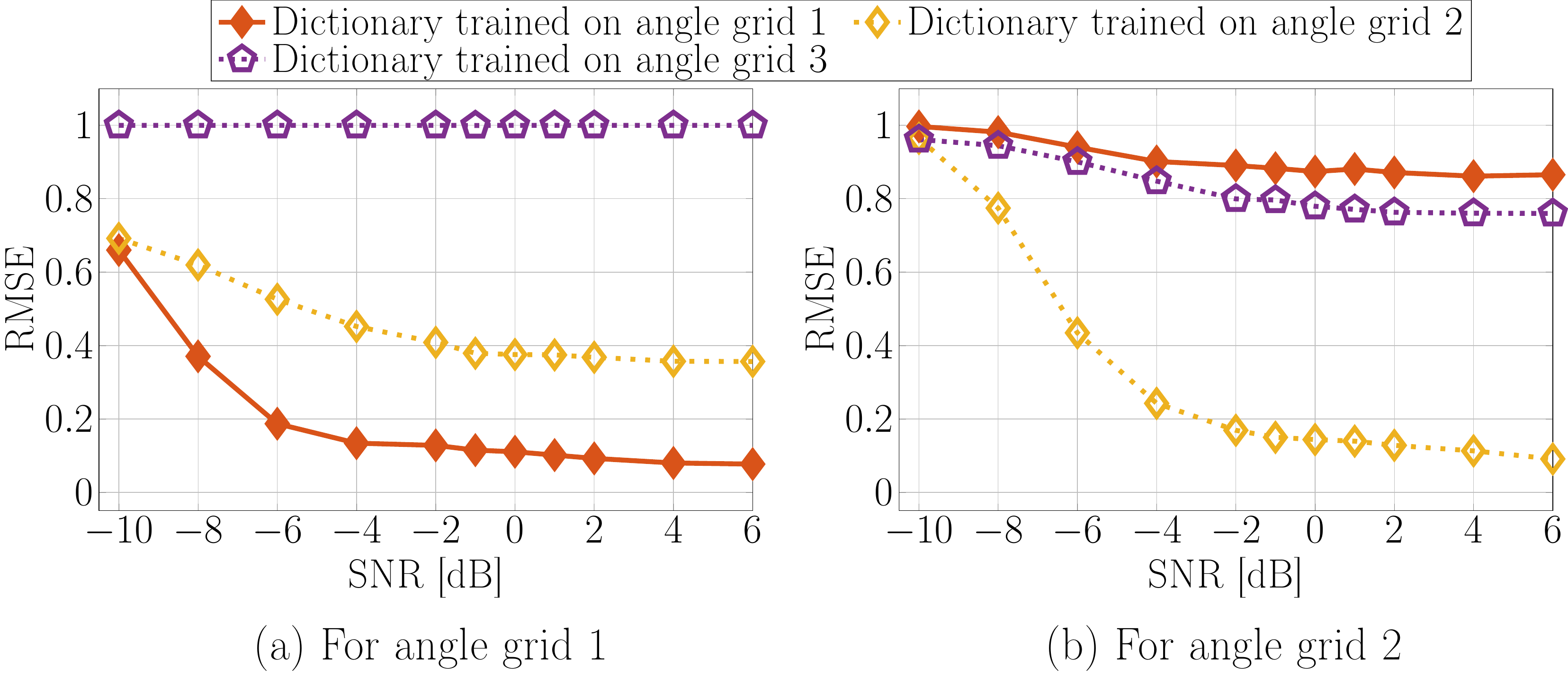}
		\caption{DoA estimation using all Dictionaries for angle grid  $1$ and $2$}
		\label{sim_figure5}
	\end{minipage}
		\begin{minipage}[t]{0.3\linewidth}
		\includegraphics[width=\textwidth]{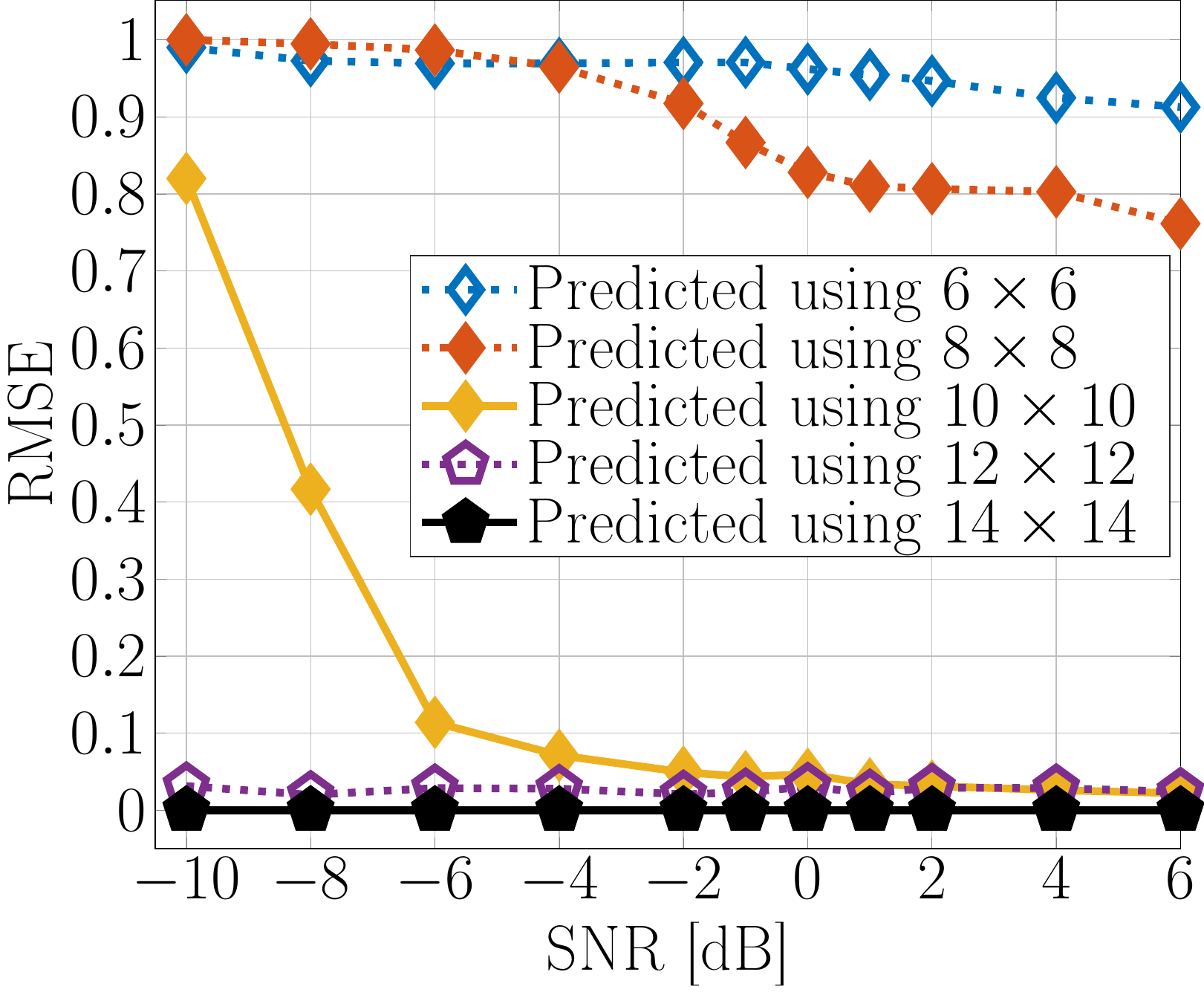}
		\caption{DoA estimation of predicted signal for $16 \times 16$ using different low antenna setups for angle grid $2$ }
		\label{sim_figure6}
	\end{minipage}%
	\vspace{-0.2cm}
\end{figure*}
\subsection{Effect of SNR in training phase to the DoA estimation of the prediction signal}  \label{snr_performace}
In this set of simulations, the performance of the algorithm is evaluated with respect to different SNR in both training and testing phases. For that purpose, the learned dictionaries are trained in the training phase using three different SNR levels. First, we consider zero noise condition, then the SNR values are set to $10$ dB and $30$ dB, respectively. In the testing phase, the SNR is changed from $-10$ dB to $0$ dB. Here, four targets ($K=4$) which are randomly located within the angle grid $2$ given in Table \ref{table_1}. DoA estimation is performed for $10000$ test cases for each SNR value in the testing phase (i.e., $[-10:0]$ dB), and the corresponding average normalized RMSE is shown in Fig. \ref{sim_figure4}. As aforementioned, in the testing phase, we use the received signal of the low antenna setup ($10 \times 10$) to predict the receive signal corresponding to the high antenna setup ($16 \times 16$). Based on the results shown in Fig. \ref{sim_figure4}, it can be seen that the RMSE of the predicted signal outperforms the RMSE of the low antenna setup in the low SNR regime. Also, it can be seen that the performance of the predicted signal is enhanced as the SNR of the training phase increases, i.e., RMSE of the predicted signal using the dictionaries trained with SNR $=30$ dB is better than the dictionaries trained with SNR $=10$ dB. However, in practice it cannot be guaranteed that higher SNR levels such as $30$ dB are available. Yet, training at SNR $=10$ dB is a possible realistic scenario, which we are going to use in the upcoming simulations. Due to space limitation, only results for angle grid $2$ is shown here. However, we observed similar behaviour for other angle grids as well.
\subsection{DoA performance of dictionaries in different angle grids} 
In these simulations, we investigate the DoA performance of the predicted signal using dictionaries which are trained on different angle grids. Here, we consider four targets randomly located within the angle grid $1$ and $2$. Here we test the feasibility of using the dictionary learned from one grid on the other grids. For instance, if the received signal of the low antenna is from the targets within angle grid $1$, then we predict the received signal corresponding to high antenna setup three times using three dictionary pairs trained using angle grid $1$, $2$ and $3$. It can be seen that the DoA estimation using the trained dictionary of the same angle range of the targets under test has the lowest RMSE for all angle ranges. Inferring which angle grid the targets belong to can be done using the low antenna setup only, then our setup is used to enhance the angular resolution.
\subsection{Effect of number of antennas in low antenna setup}  
Here, we aim at investigating the upper limit gap between the predicted high antenna setup and the actual low antenna setup. For this simulation, we consider four targets randomly in the angle grid $2$. Here, the number of antennas in the low antenna setup is changed from $6 \times 6$ to $14 \times 14 $ while the number of antennas in the predicted high antenna setup is fixed as $16 \times 16 $. Fig. \ref{sim_figure6} shows the average normalized RMSE for $5000$ test cases in each SNR value for different antenna configurations. It can be seen that, when the number of antenna is low  ($6 \times 6$ and $8 \times 8$), the DoA estimation using the predicted signal is not good. In other words, dictionary learning is unable to capture the mapping well enough due to the very large antenna gap (i.e. extrapolation of low antenna setup to high antenna setup is not in acceptable level). Note that in this case the number of antenna elements in high antenna setup is four times bigger than the number of antenna elements in low antenna setup. However, DoA estimation using the predicted signal improves when the antenna gap between the low and high antenna set up is reasonably close like $10 \times 10$ to $16 \times 16$. However, in this case number of antenna elements ratio of high antenna setup to low antenna setup is $2.56$.
\section{Conclusion} \label{con}
In this paper, we proposed a novel supervised learning algorithm based on coupled dictionary learning for antenna array extrapolation for MIMO radar using non sparse arrays. The key idea of the paper is to learn a coupled dictionary pair having the same sparse representation for the received signals of both low and high antenna setup, which can be used to map the low antenna setup to the high one. The simulations results show significantly improved performance using the predicted signal compared to the actual low antenna setup especially in noisy environments. Moreover, we performed an investigation on the upper limit on the gap between the total antennas count of the actual setup and the predicted ones. 
\bibliographystyle{IEEEbib}
\bibliography{refs.bib}
\end{document}